\documentclass[epsfig,twocolumn,floatfix]{revtex4}

\usepackage{amsmath}
\usepackage{amssymb}
\usepackage{amsfonts}

\usepackage{epsfig}
\usepackage{graphicx}

\def\la{\lambda}
\def\th{\theta}

\def\Om{\Omega}

\def\d#1#2{\displaystyle\frac{\displaystyle #1}{\displaystyle #2}}

\usepackage{color}

\definecolor{dyellow}{rgb}{1.,0.8,.0}
\definecolor{myblue}{rgb}{.1,.1,.7}
\definecolor{dcyan}{rgb}{.0,.6,.6}
\definecolor{dmagenta}{rgb}{0.6,0.0,0.6}
\definecolor{brown}{rgb}{0.6,0.2,0.}
\definecolor{darkblue}{rgb}{.0,.0,0.5}
\definecolor{darkred}{rgb}{0.75,0.0,0.0}
\definecolor{orange}{rgb}{1.,.6,.0}
\definecolor{dorange}{rgb}{0.8,.4,.0}
\definecolor{darkgreen}{rgb}{0.0,0.6,0.0}
\definecolor{purple}{rgb}{.4,.0,.4}

\def\bc{\begin{center}}
\def\ec{\end{center}}
\def\be{\begin{eqnarray}}
\def\ee{\end{eqnarray}}
\def\nno{\nonumber}
\newcommand{\omits}[1]{}
\begin{document}
\omits{\begin{CJK*}{GB}{song}
\newcommand{\song}{\CJKfamily{song}}
\newcommand{\fs}{\CJKfamily{fs}}
\newcommand{\kai}{\CJKfamily{kai}}
\newcommand{\hei}{\CJKfamily{hei}}
\CJKcaption{GB}}

%
%
%
\title{Gravitational collapse of a spherical star with heat flow\\
 as a possible energy mechanism of gamma-ray bursts\footnote{This work is partly supported by the Natural Science Foundation
of China
under Grant Nos. 90403023 and 10575106 and Knowledge Innovation Funds of CAS (KJCX3-SYW-S03).  One of the
authors (C.-B. G.) got the support in the initial stage of the present work from the Interdisciplinary Center for Theoretical Study, University of
Science and Technology of China.}}

\author{{Zhe Chang$^1$}} \email{changz@ihep.ac.cn}
\author{{Cheng-Bo Guan}$^{2}$} 
\author{{Chao-Guang Huang}$^1$} \email{huangcg@ihep.ac.cn}
\author{{Xin Li}$^1$} \email{lixin@ihep.ac.cn}
\affiliation{$^1$ Institute of High Energy Physics, Chinese Academy of
Sciences, Beijing 100049}
\affiliation{$^2$ Department of Physics, Weihai Branch of Shandong University, Weihai, Shandong, 264209}

\begin{abstract}
We investigate the gravitational collapse of a spherically
symmetric, inhomogeneous star, which is described by a perfect fluid
with heat flow and satisfies the equation of state $p=\rho/3$ at its
center. In the process of the gravitational collapsing, the energy
of the whole star is emitted into space. And the remaining spacetime
is a Minkowski one without a remnant at the end of the process.  For a star with a
solar mass and solar radius, the total energy emitted is at the
order of $10^{54}$ {\rm erg}, and the time-scale of the process is
about $8s$. These are in the typical values for a gamma-ray burst.
Thus, we suggest the gravitational collapse of a spherical star with
heat flow as a possible energy mechanism of gamma-ray bursts.

\bigskip


\bigskip


\end{abstract}

\maketitle

%

More than 30 years ago, Klebesadel, Strong and Olson announced
the discovery of the gamma-ray bursts \cite{gamma-ray}.  Since then,
thousands of gamma-ray bursts have been observed \cite{catalog}.
In particular, since the advent of
the Compton Gamma-Ray observation in 1991, the gamma-ray bursts have
been detected daily. The durations of the gamma-ray bursts range from
about 30ms to over 500s \cite{duration}. The inferred isotropic
luminosities are typically on the order of $10^{51\sim52}$ {\rm
erg/s} \cite{energy}. The abundance of observations has led to a well
described gamma-ray burst phenomenology. 

In this Letter, we study the gravitational collapse of a spherical
star with heat flow in a generalized Friedmann coordinate system.
It is supposed reasonably that a fluid star is spherically symmetric
but inhomogeneous and that the equation of
state at its center is $p=\rho/3$. We find that all energy
of the star will be emitted in the process of collapse, and an empty
flat spacetime will be left behind. For a star with about a solar mass
and a solar radius, the energy at the order of $10^{54}$ {\rm erg}
will be emitted into space within about $8s$. It is the typical value
for a gamma-ray burst in an isotropic model. Thus, we suggest the 
spherical gravitational collapse with heat flow as a possible energy 
mechanism of gamma-ray bursts, which is a different mechanism from the ones discussed 
extensively in literature \cite{model}.

In the pioneer paper to understand the late stages of stellar
evolutions \cite{OS}, the collapsing star is supposed to consist of homogeneous,
spherically symmetric, pressureless, perfect fluid and to be surrounded
by an empty space.  The interior of the star may
be described by the Friedmann-Robertson-Walker metric \cite{Wein}
\be \label{FRW}
ds^2=dt^2 -a^2(t)\left(\frac{dr^2}{1-kr^2} +
r^2d\Om^2\right),
\ee
where $d\Om^2=d\th^2+\sin^2\th d\varphi^2$ is the metric on
a unit 2-sphere, $k$ is $\frac {8\pi G}{3}$ times of the initial
energy density of dust in the unit of $c=1$, and $a(t)$ is the
solution of $\dot a^2(t)=k[a^{-1}(t)-1].$

In an astrophysical
environment, a star usually emits radiation and throws out particles
in the process of gravitational collapse.  In this situation, the
heat flow in the interior of a star should not be ignored and the
exterior spacetime is no longer described by a Schwarzschild metric.
To take the radiation of a star into account, the interior solution
of the gravitational collapse of radiating stars should match to
the exterior spacetime described by the Vaidya solution \cite{Vaidya}
\be \label{Vaidya}
ds^2=(1-\frac{2M(v)}{R})dv^2+2dv d{R}-R^2 d\Om^2 , %
\ee
which has been studied extensively \cite{BOS,more,KST}.  In particular, the
gravitational collapse of a radiating spherical star with heat flow
has been studied in an isotropic coordinate system \cite{KST}
\be
ds^2=dt^2 -B^2(t,r)(dr^2+r^2d\Om^2),
\ee
where $B(t,r)={b^2(t)}[1-\la (t) r^2]^{-1}$ with suitable $b(t)$ and
$\la(t)$ functions.  They called the solution
the Friedmann-like solution.

We study the gravitational collapse of a spherically symmetric,
inhomogeneous star in a generalized Friedmann coordinate system
\be \label{varyk}
ds^2=dt^2 -a^2(t)\left(\frac{dr^2}{1-k(t)r^2} +
r^2d\Om^2\right)
\ee
with $k$ being a function of $t$.

The stress-energy tensor of the fluid with heat flow but shear-free
and without viscosity is given by 
\be T^{\mu\nu}=
(\rho+p)u^{\mu}u^{\nu}- pg^{\mu\nu}+ q^{\mu}u^{\nu}+
q^{\nu}u^{\mu}~, %
\ee %
where $\rho$ and $p$ are the proper energy
density and pressure measured by the comoving observers
respectively, $u^{\mu}$ is the 4-velocity of the fluid, and
$q^{\mu}$ is the heat flow. $u^\mu$ and $q^\mu$ satisfy
\be
u_{\mu}u^{\mu}=1 \qquad \mbox{and} \qquad q_{\mu}u^{\mu}=0 .
\ee
For the spherically symmetric collapse, one has
\be
u^\mu = (u^0, u^1, 0,0) \quad \mbox{and} \quad q^\mu = (q^0, q^1, 0, 0).
\ee

The Einstein's field equations
\be
G_{\mu\nu}=-8\pi{G}{T_{\mu\nu}}
\ee
and the covariant conservation of stress-energy tensor give rise to
\be 8\pi{G}\rho &= & \d
{k+\dot{a}^2+2a\ddot a}{a^2}+ \d Y {a^2A^2}, \label{rho}\\
8\pi{G}p &=& - \d{k+\dot{a}^2+2a\ddot{a}}{a^2} -
           \d X {a^2A^2}, \label{eq2-2}
\ee
\be \label{u0}
u_0^{\pm}&=&\sqrt{\frac{Y (X+Y) -\frac {Z^2} 2\pm Z\sqrt{\frac {Z^2} 4- X  Y} } {(X+Y)^2 -Z^2 } }, \qquad \\
u_1&=& aA\sqrt{u_0^2-1},
\ee
\be
8\pi{G}q_0&=& \frac{1}{2a^{2}A^2}\left[\frac{Y}{u_0}-(Y-X)u_0\right],\\
8\pi{G}q_1&=& \frac{1}{2}\left[\frac{X}{u_1}-\frac {(Y-X)u_1}{a^2A^2}\right],
\ee
with the identity
\be
(8\pi G)^2 q_\mu q^\mu =- \d {Z^2-4XY}{4a^4A^4} , \label{q2}
\ee
where the dots denote the derivatives with respect to time $t$, and
\be
A&=& \frac {1} {\sqrt{1-k(t)r^2}}, \\
X&=& 3a\dot{a}A\dot{A}+a^2A\ddot{A},\\
\label{Y}%
Y&=& [2(k+\dot{a}^2-a\ddot{a})A-
     a\dot{a}\dot{A}-a^2\ddot{A}]A,\qquad \\
Z&=& -\frac{4}{r}a\dot{A}.
\ee
$X, Y, Z$ should satisfy the following solvable condition,
\be\label{Solvable}
Z^2\geq 4XY.
\ee
For spherical collapse, the dominant energy conditions \cite{Hawk} require
\be
&&\rho-p\geq 0,\label{dec2}\\
&&\rho+p\geq 2\sqrt{-q_{\mu}q^{\mu}},\label{dec1}
\ee
which lead to
\be
X+Y>|Z|\geq 0.
\ee
%

At the center of the star $r=0$,
$\dot A,\ddot A =0$ and thus $X, Z$ vanish. Hence,
\be
u_0|_{r=0}=1, \ u_1|_{r=0}=0, \ q_0^{}|_{r=0}, \ q_1^{}|_{r=0}=0, \quad
\ee
\be \label{rho0}%
8\pi{G}\rho |_{r=0}&= &3 a^{-2}{(k+\dot{a}^2)}, \\
\label{p0}
8\pi{G}p|_{r=0} &=& - a^{-2}{(k+\dot{a}^2+2a\ddot{a})} .
\ee

The exterior solution outside the collapsing star is described by Vaidya
metric (\ref{Vaidya}).
The interior solution and exterior solution should satisfy the junction
conditions on the surface of the star \cite{Santos},
\be %
&&R_s=ar_s~,\label{vaidya-r}\\
&&\left . \frac{dr}{dt}\right |_s= \left .-a^{-2}A^{-2}\frac{u_1}{u_0}\right |_s,\label{rdot}\\
&&p^2_s=|q_\mu q^\mu|_s, \label{pq} %
\ee
and
\be
&&\left [dt^2-a^2A^2dr^2 \right ]_s \nno \\
&=& \left [\left(1-\d {2GM(v)}{R} \right)
dv^2+2 dv dR \right ]_s.
\ee
The Misner-Sharp mass \cite{Misner-Sharp} for the star is then
\be
M(v)=
\left . \frac{a}{2G}(k+\dot{a}^2)r^3 \right |_s~.\label{vaidya-m} \ee

\begin{figure}[ht]
\includegraphics[scale=0.72]{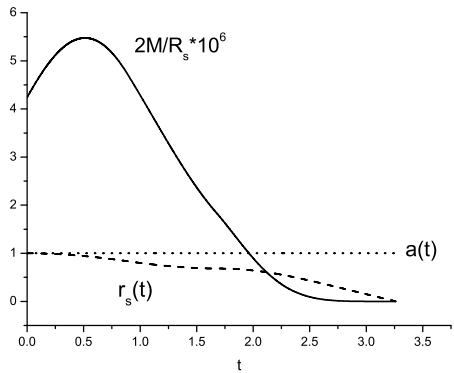}
\caption{Numerical solution of the gravitational collapse of
spherical star with heat flow in the case $p|_{r=0}={\rho}/3|_{r=0}$
and the $u_0^-$ solution in Eq.(\ref{u0}). The horizontal axis is
the time $t$ in the unit of $R_0/c$. In the process of evolution,
$r_s(t)$ (dash curve, in the unit of $R_0$) decreases monotonically
while $a(t)$ (dotted curve) keeps almost a constant.  $2M/R_s$ (real
curve) in the unit of $c^2/G$ increases first and then goes to 0 as
$R_s \to 0$, which implies that the star disappears at the end of
the collapse.}
\end{figure}
To make further
calculations, the equation of state for the matter of a star at
$r=0$ needs to be assigned.  Without loss of generality, we choose
$a(0)=1$. After the equation of state is assigned, one may obtain
the expression for $a(t)$, $k(t)$, $\rho(r,t)$, $p(r,t)$ and
$r_s(t)$ by solving Eqs.(\ref{rho0}), (\ref{p0}), (\ref{rho}),
(\ref{eq2-2}), (\ref{q2}), (\ref{rdot}), (\ref{pq}).  Then, one may
determine the Misner-Sharp mass $M(v)$ and the surface radius $R_s$
from Eqs.(\ref{vaidya-m}) and (\ref{vaidya-r}).   Fig. 1 presents
the numerical solution in the case $p|_{r=0}={\rho}/3|_{r=0}$ and
the $u_0^-$ solution in Eq.(\ref{u0}) with the initial conditions
$r_s(t=0)=R_s(t=0)=R_0$, $\dot a(t=0)=0$, $\dot R_s(t=0)\approx 0$,
which means the surface of star is almost stationary in the initial
state. Since $\rho$ and $p$ are determined by Eqs.(\ref{rho}) and
(\ref{eq2-2}), the initial state of the star is not homogeneous and
the equation of state in the star is generally deviated from the
radiation.  In the figure, the time and radius $r_s$ are in the
units of $R_0/c$ and $R_0$, respectively. In the process of
evolution, $r_s(t)$ (dash curve) decreases monotonically while
$a(t)$ (dotted curve) almost keeps a constant.  $2M/R_s$ (real
curve) in the unit of $c^2/G$ increases first and then goes to 0
more quickly than $R_s$ itself as $R_s \to 0$.  It implies that the
star will disappear without the appearance of a horizon in the
process, leaving Minkowski spacetime behind.

\begin{figure}[b]
\includegraphics[scale=0.72]{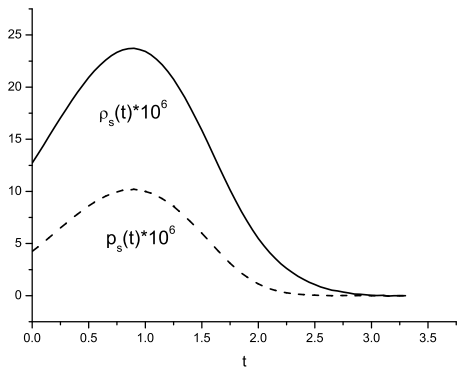}
\caption{Evolution of $\rho$ and $p$ at the boundary.  The
horizontal axis is the time in the unit of $R_0/c$.  $\rho$ and $p$
are in the unit of $c^4/(8\pi G R_0^2)$. }
\end{figure}
Fig. 2 shows the evolution of $\rho$ and $p$ at the boundary, in unit of $c^4/(8\pi GR_0^2)$.  The emission of the star arrives its maximum value at about $0.9 R_0/c$ after the beginning of the collapse. In the late stage of the process, the equation of state at the boundary as
well as at each point in the star tends to $p=\frac 1 3 \rho$.  At the end
of the process both the energy density and the pressure become
0, which confirms that the whole star is radiated out into space
in the process.

Furthermore, our numerical analysis shows that the star with $p|_{r=0}=\frac 1 3 \rho|_{r=0}$
will radiate its whole mass in the process of the collapse, without the
appearance of a horizon, for different initial values.

In conclusion, we have presented a solution of the Einstein equations, which describes 
the gravitational collapse of an inhomogeneously-distributed but 
spherically-symmetric star consisting of perfect fluid with heat flow 
but without viscosity.  We have shown that if the star satisfies 
Eqs.(\ref{rho}) and (\ref{eq2-2}) and $p=\rho/3$ at the center, the 
whole star will be emitted in the process of the gravitational collapse 
in about $3.3R_0/c$ interval, without the formation of a horizon,
and a Minkowski spacetime is left at the end of the process.  
For a star with about a solar mass $M_\odot$ and a solar radius 
$R_\odot$, a huge mount of energy (about $1.8\times 10^{54}$ erg) 
will be emitted into space within 7.66 seconds, which are typical 
values for a gamma-ray burst.  Thus, the gravitational collapse of 
such a star might provide an energy mechanism for gamma-ray bursts.

Since heat is flowing across the boundary of the collapsing 
sphere and the pressure of the fluid at the boundary is not 
zero, the exterior space cannot be considered as vacuum. We 
have adopted a Vaidya space as the exterior of the spherical 
collapsing stars. In the Vaidya space, massless radiation 
is dominated. The spectra of the radiation from collapsing 
stars depend on the equation of state of interior 
matter. We wish the combination of the model with 
phenomenological studies on possible evolution process of the 
equation of matter state can give a good fit to the spectra 
of the gamma-ray bursts.

As a solution of Einstein equations, the advantage of the model is that it transmits all initial 
energy of a star into radiation and then supplies up to 
$10^{54}$ erg radiated energy by a popular star with solar mass. 
In this collapsing scenario, the gamma-ray bursts are supposed to 
be spherical.  Thus, one need not try to form beaming of energy, 
whose dynamical mechanism is not well understood up to now.  
In fact, the most important motivation of the 
study is try to find a possible candidate for the replacement 
of the models for the gamma-ray bursts including beaming.  Of course, other 
ways should not be ignored.   

In the model, the initial energy, which is transformed into 
radiation and emitted completely in the process, and initial radius 
of a star are only parameters of the solutions.  Different initial
energy and radius will result in different energy release and
different collapsing timescale.  Therefore, the model for the 
energy mechanism of gamma-ray burtsts will supply a wide spread
in total energy release and various timescales.  

Finally, it should be stressed that if the solution really serves 
as a model for gamma-ray bursts, further studies on many problems are 
needed, which have been out of the coverage of the letter, such as, 
the trigger of collapse, the mechanism of energy 
transformation, the constraints on the spread in total energy release and timescale in the observation, light curve, spectrum of emmission, and so on.

\medskip

We would like to thank Professors H.-Y. Guo and Z.-J. Shang for helpful discussion. 

%

\end{document}